\documentclass[prl,superscriptaddress]{revtex4}

\usepackage{graphicx}
\usepackage{dcolumn}
\usepackage{amsfonts}

\usepackage[pdftex,bookmarks,colorlinks,breaklinks]{hyperref}  
\hypersetup{linkcolor=red,citecolor=blue,filecolor=pink,urlcolor=magenta} 

\begin{document}

\preprint{}

\title{Nonlinear relaxation of 0-dimension-trapped microcavity polaritons}
\author{Ounsi El Daif}
\affiliation{Institut des nanotechnologies de Lyon (INL), UMR CNRS 5270, Ecole centrale de Lyon, 36 Avenue Guy de Collongue, 69134 Ecully Cedex, France}

\author{Ga\"el Nardin, Taofiq K. Para\"{\i}so, Augustin Baas, Maxime Richard, J.-P. Brantut*, Francois Morier-Genoud, Benoit Deveaud-Pl\'edran}
\affiliation{Institut de Photonique et d'{\'E}lectronique Quantiques, {\'E}cole Polytechnique F{\'e}d{\'e}rale de Lausanne (EPFL), CH-1015 Lausanne,
Switzerland. *Now at Laboratoire Charles Fabry de l'Institut d'Optique, Campus polytechnique, RD128, 91127 PALAISEAU FRANCE}

\author{Thierry Guillet}
\affiliation{Groupe d'\'{e}tude des semiconducteurs (GES), Universit{\'e} de Montpellier II and CNRS, Place Eug{\`e}ne Bataillon, F-34095 Montpellier cedex 5, France}

\begin{abstract}

We study the emission properties of confined polariton states in shallow zero-dimensional traps under non-resonant excitation. We evidence several relaxation regimes. For slightly negative photon-exciton detuning, we observe a nonlinear increase of the emission intensity, characteristic of carrier-carrier scattering assisted relaxation under strong-coupling regime. This demonstrates the efficient relaxation towards a confined state of the system. For slightly positive detuning, we observe the transition from strong to weak coupling regime and then to single-mode lasing. 
\end{abstract}


\maketitle

Research on solid semiconductor systems has been motivated, during the last 50 years, by both applied and basic goals. From a fundamental point of view, solid state systems were expected to show various effects, ranging from the Purcell effect \cite{purcell} to full quantum confinement. Added to the latter, Bose-Einstein condensation is expected at rather high temperatures. One of the main fields of research in semiconductor physics has therefore been the quest for a possible Bose-Eintein condensation (BEC) of excitons, initially proposed by Moskalenko \cite{moskalenko,blatt} in 1962. A clear demonstration of pure excitonic BEC has not been achieved yet, presumably because of several competing effects such as disorder and Auger effect \cite{ohara}. Present research directions aim at the realization of a trap for excitons \cite{butov02b}. 
Recently, BEC has been achieved for microcavity polaritons in a CdTe based microcavity \cite{BECpol}, and in GaAs with a trap \cite{snokebec}. 
Indeed, these quasi-particles have the great  advantage over excitons or electrons to exhibit a very light effective mass, thanks to their photon component. BEC was favored by a confinement of the polaritons within small volumes \cite{BECpol} in local defects of the microcavity, an aspect confirmed by theoretical works \cite{sarchi}. The population of the ground state of the system in these defects was favored by Coulomb interaction, this is shown by the presence of a density dependence of the relaxation towards the ground state of the system, before the stimulation characteristic of condensation. It was already observed and predicted in several experimental and theoretical works: before condensing and showing a stimulation of the population of the ground (or condensed) state of the system, polaritons are expected to relax linearly (through phonons) and then quadratically, through Coulomb interaction \cite{senellart,porras02} towards low energy states.

At the same time, most breakthroughs in semiconductor physics and technology over  the last thirty years originated from quantum confinement of elementary excitations along one, two, or three spatial dimensions \cite{hess,bimberg} and from the improvement of their coupling to the electromagnetic field. Indeed, confinement in semiconductor  structures allows design and shaping of their electrical and optical properties. 
Such confinement allows to control the emission properties of matter, and can be used for applications in many fields, ranging from optoelectronics to quantum information. 

We present here studies under non-resonant excitation of trapped polaritons: we performed detailed studies of the behavior of the photoluminescence (PL) emission of the ground state of the confined system, while increasing the pump power (the density of created population). In order to understand the role of the various relaxation channels, nonlinearities are identified, and the coupling regime (weak or strong) is carefully followed.

\begin{figure}[h]
   \centering
   \includegraphics[width=0.8\textwidth]{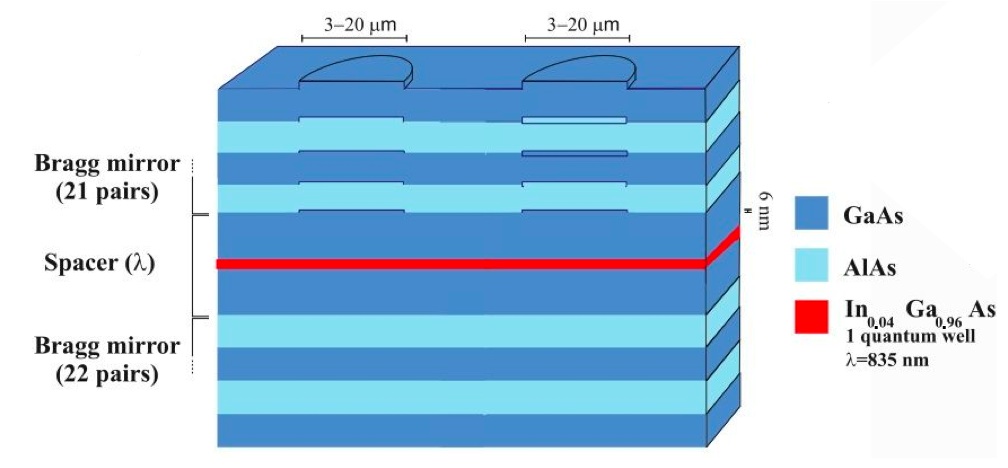} 
   \caption{(a) Scheme of the sample.}
   \label{images1}
\end{figure}

We fabricated a GaAs/AlAs two-dimensional (2D) semiconductor microcavity with an embedded InGaAs quantum well, giving rise to a Rabi splitting of $3.5 meV$. We have designed and realized $6 nm$ high mesas of $3 - 19\mu m$ in diameter on the surface of the spacer layer, which act as zero-dimensional (0D) traps for the photons, a drawing of the structure is shown on Fig. \ref{images1}. We previously demonstrated and characterized strong-coupling in 0 and 2 dimensions for the various mesa sizes \cite{mathese,nousAPL,nousPRB,nousPSS}. The 0D structure shows a very high-Q factor of $\sim 21000$, thanks to the fact that the confining mesas have sizes smaller than the typical optical disorder in GaAs microcavities \cite{mathese,sanvitto,savona}.

\begin{figure}[h]
   \centering
   \includegraphics[width=0.8\textwidth]{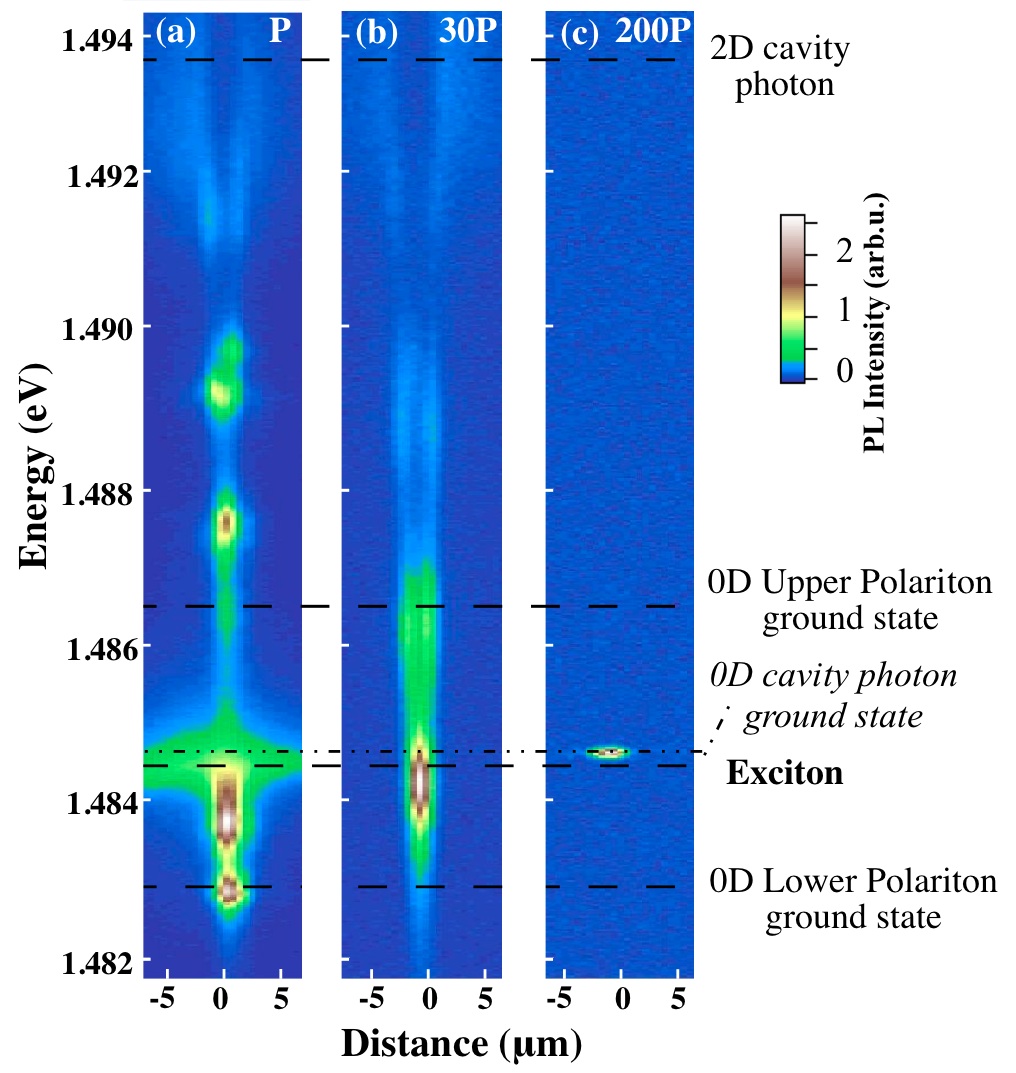} 
   \caption{The series of three images shows the photoluminescence (PL) signal of a $15\mu m$ long stripe of the sample's surface, under non-resonant excitation on a mesa showing $\delta_{0D} \simeq 0.2 meV$. Various pump powers (and thus regimes) are shown: (a) strong coupling regime under a low pump power (b) towards weak coupling (c) lasing on the 0D photon ground state indicated by the point-dashed line.}
   \label{images2}
\end{figure}

The spectral properties and PL emission intensity of the sample are measured. The
sample, cooled to about 10K, is pumped by a TiSa laser ($\lambda \approx 780 nm$), under increasing pump power. The luminescence is analyzed with a $25\mu eV$ resolution spectrometer followed by a nitrogen cooled CCD. The experimental set-up allows to image the surface of the sample on the slit of the spectrometer, and thus to perform spatially and spectrally resolved images. The laser spot has a diameter of $15 \mu m$ and each measurement is performed on a zone containing one $3 \mu m$-diameter mesa.

Various mesas were studied, we focus here on two different, showing a detuning ($\delta = E_{photon}-E_{exciton}$) respectively of $\delta_{0D} \simeq 0.2 meV$ and $\delta_{0D} \simeq -1meV$. $\delta_{0D}$ being defined as the energy difference between the uncoupled ground photon state and the exciton. This allows to work with various exciton component for the confined polaritons.

Fig. \ref{images2} shows the luminescence spectra of the polariton energy along one dimension of the plane of the cavity, under low excitation power. This luminescence can be related to the modes' squared wave functions in real space, although the intensities at various energies should not be compared as they are modulated by the relaxation. Added to that, it must be paid attention to the cylindrical asymmetry of the mesas, which do not allow to reconstruct a 2D spatial image from one dimension. One can see the 2D signal at $\approx 1.4846 eV$, corresponding to a 2D lower polariton at the exciton's energy, because the 2D system is at positive detuning ($E_{photon}>>E_{exciton}$) on this range of positions, the upper 2D polariton can also be seen at the 2D cavity mode's energy, at $\approx 1.4936 eV$. On a distance of $3 \mu m$ (ie within the mesa) in the center of the image, two confined states can be seen for the 0D lower polariton, and four for the 0D upper polariton. The confined ground state is at slightly positive detuning. One can see lobes on the excited modes of the upper 0D polariton.

Despite an excitation spot larger than the confining structure, the relaxation towards 0D states occurs. The spot excites at high energy a 2D electron-hole plasma, which is known to relax towards the 2D polariton states \cite{porras02}. Fig. \ref{images}(b) shows that relaxation towards 0D states occurs also. It is nevertheless not possible at this point to determine the relaxation channel: directly from the 2D plasma to the confined state, or through the 2D polariton states. Time-resolved studies are being performed to understand further the various relaxation mechanisms \cite{tao}. In any case the 0D states are populated from 2D states or continuum, allowing to identify the 0D structures as efficient traps, which is a first important feature for a possible condensation.

\begin{figure}[h]
   \centering
   \includegraphics[width=0.75\textwidth]{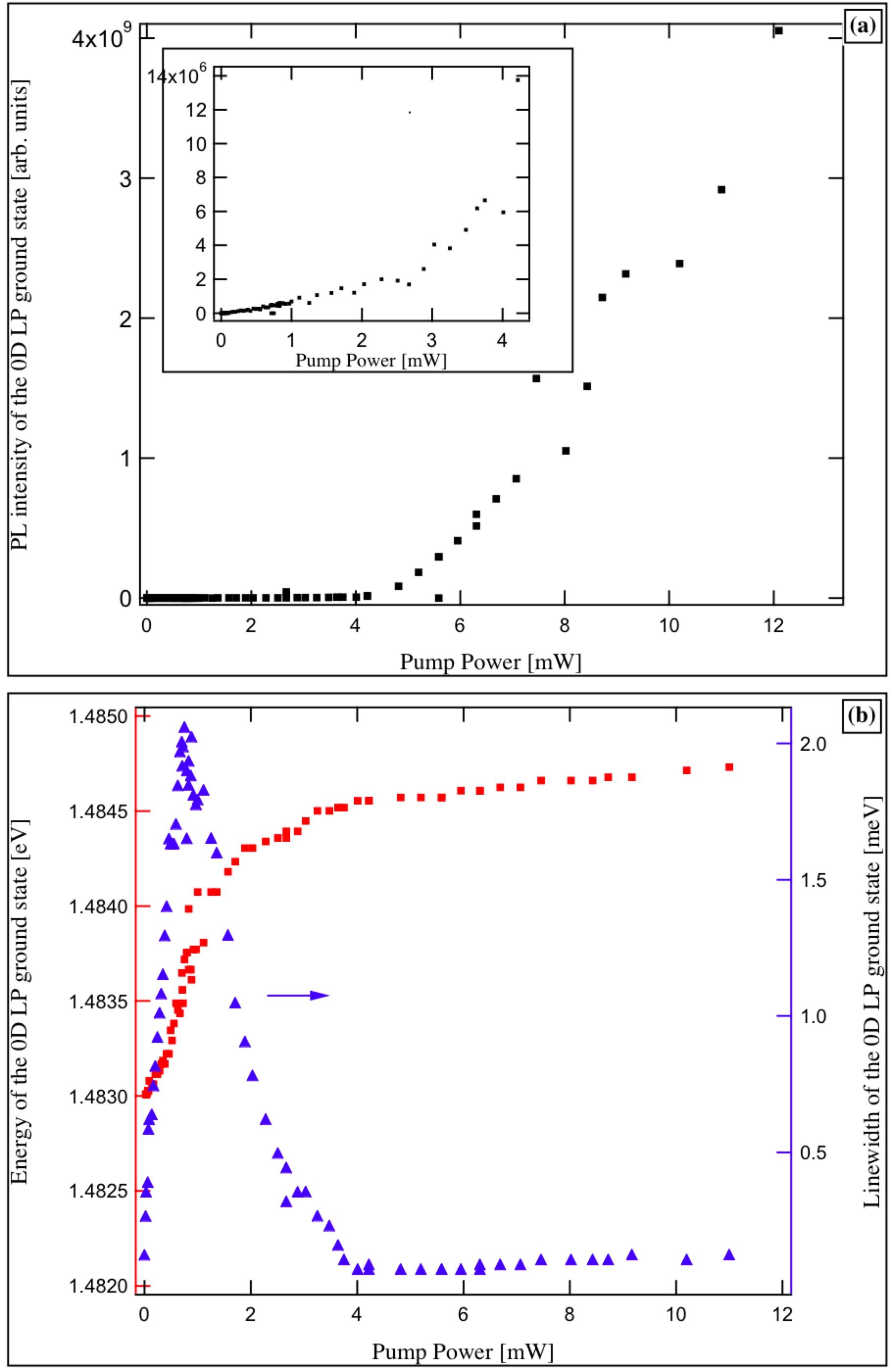} 
   \caption{PL intensity (a) and linewidth and blueshift (b) of the confined ground mode luminescence peak at a slightly positive detuning $\delta _{0D ground state}\simeq 0.2 meV$, the inset in (a) is a zoom on the first nonlinear threshold.}
   \label{fig2}
\end{figure}

\begin{figure}[h]
     \centering
   \includegraphics[width=0.75\textwidth]{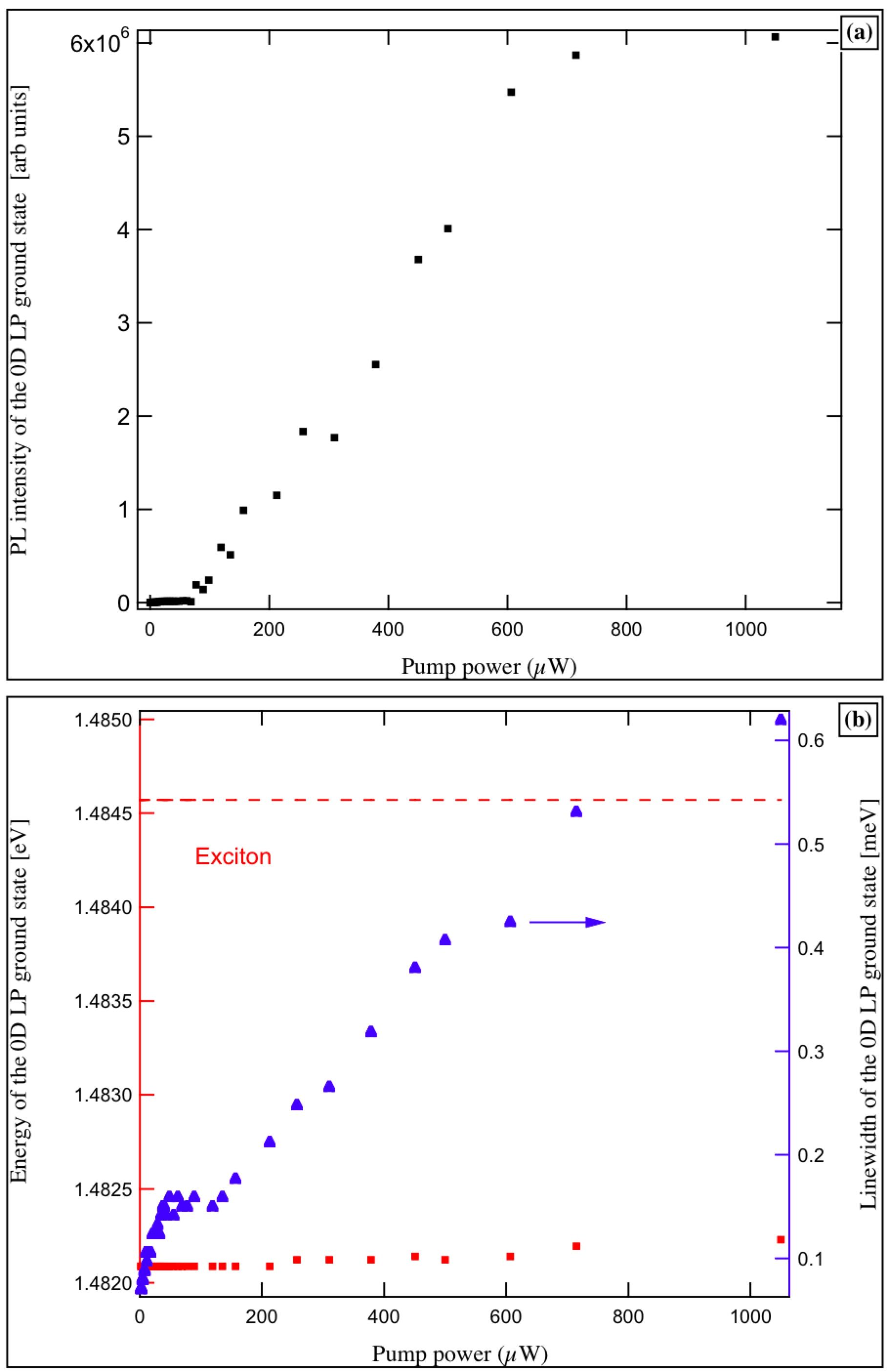} 
   \caption{PL intensity (a) and linewidth and blueshift (b) of the confined ground mode luminescence peak at negative detuning $\delta _{0D ground state}\simeq -1 meV$.}
   \label{fig3}
\end{figure}

Figs. \ref{fig2} and \ref{fig3} show the PL emission characteristics of the 0D ground state as a function of the pump power, for two different detunings. Before discussing into detail each situation, let us, for both of them, see the general trends. Indeed, for both cases, several similar behaviors can clearly be distinguished, the first one is a linear evolution of the photoluminescence (PL) emission intensity, due to energy relaxation towards the polariton states through phonons. It is followed by a nonlinearity, that we will further discuss. Due to the further pump power increase, the excitons density increases and crosses progressively the Mott \cite{kappei} saturation density. This yields a progressive loss of the strong-coupling regime. This loss of the strong-coupling can be assessed by the comparison between the polariton's energy and the uncoupled photon's energy. This loss occurs under different conditions depending on $\delta_{0D}$. Let us now analyze precisely the regimes, depending on this detuning.

In the case of Fig. \ref{fig2}(a) and (b), at a slight positive detuning of $\delta _{0D}\simeq 0.2 meV$, the first nonlinearity occurs at a pump power of $3 mW$ (as seen in the inset of Fig. \ref{fig2}(a)) while the strong coupling is already lost in the system, indeed the strong excitonic component of the studied state yields an important increase in the linewidth while increasing the density. Therefore the loss of the strong-coupling can already be assessed at a pump power of $1 mW$, given the fact that the luminescence peak, due to its blueshift and broadening, crosses the uncoupled exciton's energy. This situation is illustrated by the succession of images in the real space, Fig. \ref{images2}(a) was taken at the first nonlinear threshold and illustrates the transition towards weak coupling. (We also reported the observation of this strong-to-weak cross-over for a different kind of mesa \cite{nousPSS} through reciprocal space imaging.) In weak coupling regime, after the nonlinear increase, a second nonlinearity occurs: an exponential increase of the emission at a pump power of $\simeq 4 mW$ (Fig. \ref{fig2}(a)) along with a sharpening of the linewidth (Fig. \ref{fig2}(b)) characteristic of a laser -VCSEL- it is nicely illustrated by Fig. \ref{images2}(b) where only the ground state emission is visible at the uncoupled confined ground photon mode energy. 

In opposition, at a negative detuning of $\delta _{0D}\simeq -1 meV$, as shown on Fig. \ref{fig3}(a) and (b), the nonlinearity clearly occurs in strong-coupling regime: the emission intensity is first linear. Then, at a pump power of $100 \mu W$ a nonlinearity and a "breaking" of the slope occurs. The ground state is populated more efficiently while the density increases. As the latter favors polariton-free carriers interaction, and exciton-exciton scattering in the exciton reservoir at high \textbf{k} \cite{porras02}, we can deduce that these effects, due to Coulomb interaction, are now dominating and populating the system's ground state. Therefore it is possible to assess that the polariton traps can be populated thanks to Coulomb interaction effects. 

To sum up we can extract the main points from these observations as following: we observed nonlinear behavior of trapped (0D) polaritons, and were able to show the role of Coulomb interaction on the relaxation from 2D high energy states towards the ground state of 0D polaritons. We also showed that for positive detuning, the nonlinearities occur in a regime of weak-coupling, and that it is possible to observe a VCSEL behavior on the ground confined photon mode of the weakly coupled system. Finally we observed that the nonlinearities occur in strong coupling at slightly negative detuning, where the the excitonic weight of the polaritons is optimal: low enough to ensure strong coupling at high densities, high enough to ensure Coulomb interaction. This result is consistent (although the excitation conditions differ) with experiments performed with an excitation resonant with the lower polariton branch where the lowest threshold for nonlinear effects appears at a negative detuning, see ref\cite{kund04} and references therein.

To conclude, although we did not observe any condensation effect in strong-coupling, we evidenced nonlinear behaviors characteristic of scattering effects. The fact that in precise conditions, the threshold of this nonlinear regime appears after the loss of the strong-coupling, allows strong hope towards possible condensation in similar samples. The latter need to include a higher number of quantum wells, in order to increase the weak-coupling threshold for a given density of excitations in the system. We also expect parametric effects in between confined states and between confined and extended states, and the observation of coherence transfer between states or coherent effects could be made \cite{kund04}. As all the measurements shown here were performed under non-resonant excitation, it is possible, thanks to the recent observation of a microcavity light emitting diode \cite{bajoni08}, to envision electrical injection. One can therefore already think about optoelectronics applications (as a quasi-thresholdless single mode laser) through its combination with future coherent effects in the polariton traps presented here. 

\

We acknowledge financial support from the Swiss National Centre for Competence in Research (NCCR) Quantum Photonics. We would like to thank Davide Sarchi, Stefano Portolan and Vincenzo Savona for fruitful discussions.


\newpage

\end{document}